\documentclass[useAMS,usenatbib]{mn2e}

\usepackage{graphicx}
\usepackage{amssymb}
\usepackage{epsfig}
\usepackage{amssymb}
\usepackage{graphicx}
\usepackage{commath}

\title[]{The ALMA early science view of FUor/EXor objects - V:  continuum disc masses and sizes}
\author[L. Cieza et al.]{
\Large
Lucas A. Cieza$^{1,2}$, 
Dary Ru\'iz-Rodr\'iguez${^3}$,
Sebastian Perez$^{2, 4}$,
Simon Casassus$^{2, 4}$, 
\noindent \vspace{0.2cm} \\
\Large \rm Jonathan P. Williams$^{5}$, 
Alice Zurlo$^{1, 2}$,
David A. Principe$^{6}$,
Antonio Hales$^{7,8}$,
Jose L. Prieto$^{1,9}$, 
\noindent \vspace{0.2cm} \\
\Large \rm John J. Tobin$^{10,11}$, Zhaohuan  Zhu$^{12}$, Sebastian Marino$^{13}$
\vspace{0.2cm} \\
$^{1}$N\'ucleo de Astronom\'ia,  Facultad de Ingenier{\'i}a y Ciencias, Universidad Diego Portales,  Av. Ejercito 441, Santiago, Chile\\
$^{2}$Millenium Nucleus ``Protoplanetary discs in ALMA Early Science",  Universidad Diego Portales,  Av. Ejercito 441, Santiago, Chile \\
$^{3}$Research School of Astronomy and Astrophysics, Australian National University, Canberra, ACT 2611, Australia\\
$^{4}$Departamento de Astronom\'ia, Universidad de Chile, Casilla 36-D Santiago, Chile\\
$^{5}$Institute for Astronomy, University of Hawaii at Manoa, Honolulu, HI 96822, USA\\
$^{6}$Massachusetts Institute of Technology, Kavli Institute for Astrophysics, Cambridge, MA, USA\\
$^{7}$Joint ALMA Observatory, Alonso de Cordova 3107, Vitacura 763-0355, Santiago - Chile\\
$^{8}$National Radio Astronomy Observatory, 520 Edgemont Road, Charlottesville, Virginia, 22903-2475,  USA \\
$^{9}$Millennium Institute of Astrophysics, Santiago, Chile\\
$^{10}$Homer L. Dodge Department of Physics and Astronomy, University of Oklahoma, 440 W. Brooks Street, Norman, Oklahoma 73019\\
$^{11}$Leiden Observatory, Leiden University, P.O. Box 9513, 2300-RA Leiden, The Netherlands \\ 
$^{12}$Department of Physics and Astronomy, University of Nevada, Las Vegas, 4505 South Maryland Parkway, Las Vegas, NV 89154, USA\\
$^{13}$Institute of Astronomy, University of Cambridge, Madingley Road, Cambridge CB3 0HA, UK
}
\begin{document}



\maketitle

\label{firstpage}

\begin{abstract}
\noindent 
Low-mass stars build a significant fraction of their total mass during short outbursts of enhanced accretion known as FUor and EXor outbursts.  FUor objects are characterized by a sudden brightening of $\sim$5 magnitudes at visible wavelengths within one year and remain bright for decades.  EXor objects have lower amplitude outbursts on shorter timescales.  Here we discuss a 1.3 mm ALMA  mini-survey of eight outbursting sources (three FUor, four EXor, and the borderline object V1647 Ori) in the Orion Molecular Cloud. While previous papers in this series discuss the remarkable molecular outflows observed in the three FUor objects and V1647 Ori, here we focus on the continuum data and the differences and similarities between the FUor and EXor populations. We find that FUor discs are significantly more massive ($\sim$80-600 M$_{JUP}$) than the EXor objects ($\sim$0.5-40 M$_{JUP}$). We also report that the EXor sources lack the prominent outflows seen in the FUor population. Even though our sample is small, the large differences in disc masses and outflow activity suggest that the two types of objects represent different evolutionary stages.The FUor sources seem to be rather compact (R$_{c}$ $<$ 20-40 au) and to have a smaller characteristic radius for a given disc mass when compared to T Tauri stars. V1118~Ori, the only known close binary system in our sample, is shown to host a disc around each one of the stellar components. The disc around HBC~494 is asymmetric, hinting at a structure in the outer disc or the presence of a second disc.
\end{abstract}



\section{Introduction}

The observed protostellar luminosities are typically significantly lower than the luminosity expected from steady protostellar disc accretion to build $\sim$1 M$_{\odot}$ in $\sim$1 Myr. This was first discovered by Kenyon et al. (1990) but drew great attention in the star formation community due to its confirmation by the Spitzer Cores to discs survey (Evans et al. 2009). One solution to this  ``Luminosity Problem" is that protostellar discs accrete episodically, 
spending most of their evolutionary time accreting slowly with low luminosities but undergoing outbursts of rapid accretion during which most of the stellar mass is accreted. 
This episodic accretion picture challenges the traditional steady accretion model and may significantly alter our understanding of star and planet formation (Zhu et al. 2009, Dunham $\&$ Vorobyov 2012, Cieza et al. 2016), binary formation (Stamatellos et al.  2012), the luminosity spread in young clusters (Baraffe et al.  2012, Hosokawa et al. 2011),  disc chemistry (Visser $\&$ Bergin 2012), and the surrounding envelope (Jorgensen et al. 2015). 
Understanding the origin of disc episodic accretion is thus crucial for both star and planet formation theory. 
The most extreme episodic accretion events in YSOs are FUor and EXor outbursts, 
named after the prototypes FU Orionis and EX Lupi, respectively. 
FUor objects suddenly brighten by 5 magnitudes or more within one year (Herbig 1977) and remain bright for decades, although most of them have been discovered at  the high-luminosity state (have not been seen erupting) and are technically classified as FUor-like objects based on their spectra. 
EXor objects have shorter and lower amplitude outbursts (Herbig 2007; Audard et al. 2014), although the physical distinctions between FUor and EXor outburst remain unclear.
The sudden brightness enhancement is due to an abrupt accretion rate increase  (\.M reaches up to 10$^{-4}$ M$_\odot$/yr, Hartmann $\&$ Kenyon 1996). 
Hartmann (1998) argues that disc accretion is inherently intermittent and  that the main difference between FUor and EXor outbursts is the evolutionary stage in which they take place, with 
FU Ori outburst occurring preferentially during the embedded stage and EXor outbursts occurring during the T Tauri phase. 
At least four outburst mechanisms have been proposed to date: 1) the coupling of magnetorotational and gravitational instabilities (MRI+GI, Armitage et al. 2001, Zhu et al. 2009, Martin et al. 2012),  2) disc fragmentations followed by the inward migration of the resulting fragments (Vorobyov $\&$ Basu 2005, Zhu et al. 2012), 
3) Thermal-viscous instability (Bell et al. 1995),   and 4)  instabilities  induced by planets (Clarke et al. 1990, Lodato $\&$ Clarke 2004) or stellar companions (Bonnell $\&$ Bastien 1992).

While many FUor objects have been studied by (sub)millimeter observations (e.g., Polomski et al. 2005, Perez et al. 2010, Dunham et al. 2012, Fischer et al. 2012, Hales et al. 2015, Cieza et al. 2016), 
resulting in estimates of masses, sizes, and/or surface density profiles of their discs,  observations of EXor objects have been rare
(Andrews et al. 2004, Liu et al. 2016).
The boundary between FUor objects and EXor objects remains vague:
After its initial outburst in 2010, HBC 772 quickly faded, similar to EXor objects. However, before
returning to the quiescent state, 
it gradually becomes bright again starting in 2011 and  roughly remains the same brightness
since  2013 (K{\'o}sp{\'a}l et al. 2016). The long duration of the outburst since then resembles the behavior of FUor objects.
Similarly, V1647 Ori has been classified as both and EXor and FUor object  (Audard et al.  2014), suggesting that these two groups of outbursts may have similar origins.  
Here we present results from a small ALMA 230 GHz (band-6)  survey of  eight outbursting sources (three FUors, four EXors, and the borderline object V1647 Ori) in Orion, studying them as a whole sample in order to establish differences and similarities between the two types of objects.

\section{Sample selection and ALMA Observations}

\subsection{Sample Selection}

All FUors and EXors in our sample were selected from the review article by Audard et al. (2014) except for ASASSN--13db.
This source was identified as a new EXor in outburst in 2013 by the All-Sky Automated Survey for Supernovae (ASAS-SN; Shappee et al. 2014) and studied in detail in Holoien et al. (2014) and Sicilia-Aguilar et al. (2017). ASASSN-13db is the lowest mass star known to experience accretion outbursts, with a spectral type of an M5 T-Tauri star in quiescence (Holoien et al. 2014), and it had a long second outburst that lasted three years (2014-2017; Sicilia-Aguilar et al. 2017).
We note that FU Ori itself is not part of our survey as it is located $>$ 10 degrees from the rest of the targets and hence could not be grouped 
in the same Observing Goal to share phase calibrators.  ALMA 350 GHz (band-7) observations of FU Ori itself are presented by Hales et al. (2015).
Our target list is shown in Table~1 and includes three FUor (V883 Ori, HBC 494, and V2775 Ori), four  EXors (NY Ori, V1143 Ori, V1118 Ori,  and ASASSN--13db), and
V1647 Ori, which can be considered an intermediate case between FUor and EXor objects (Aspin et al. 2006; Aspin 2011). 
Previous papers from this series discuss the prominent molecular outflows of the FUor tagets:  V2775 Ori  (Zurlo et al. 2017; Paper I), HBC 494 (Ruiz-Rodriguez et al. 2017a; Paper II), and V883 Ori (Ruiz-Rodriguez et al. 2017b; Paper III),  and of V1647 Ori (Principe et al. in press;  Paper IV).

\subsection{Observations}

Our band-6 observations were taken under ALMA program 2013.1.00710.S on three different dates.
Two of three observations took place on December 12$^{th}$, 2014 and April 5$^{th}$, 2015. The
precipitable water vapor (PWV) was measured at 0.7 and 1.3mm for the December and April observations, respectively.
The configuration of ALMA for both observing runs was with 34 antennas (12-m
of diameter) with baselines ranging from 14.6 to 348.5m.
The third observations were on August 30th, 2015 with an array of 35 antennas and longer baselines of
42-1574m.
The PWV during these observations was 1 mm.
The data set combining the two array configurations reached a resolution of  0.25$''$ while maintaining a  maximum recoverable
scale of 11$''$. 
In all cases,  the ALMA correlator was configured so that three spectral windows with 58.6 GHz bandwidths were centered at at 230.5380, 220.3987, and  219.5603 GHz to cover the  $^{12}$CO J = 2-1,   $^{13}$CO J = 2-1, and  C$^{18}$O J = 2-1 transitions, respectively.  
The first spectral window has a 0.04 km s$^{-1}$ spectral resolution, while the other two have a 0.08 km s$^{-1}$ resolution. 
Two additional spectral windows with 1.875 GHz bandwidths were centered at 232.6 and  218.0 GHz and were selected for continuum observations.  
Ganymede was used as  flux calibrator, while the quasars
J0538-4405 and J0541-0541 were observed for bandpass and phase
calibration respectively.
Observations of the phase calibrator were alternated with the
science target every 5 minutes to calibrate the time dependence variations of the complex
gains. The total integration time was 122 s per science target.

\subsection{Data reduction}

All data were calibrated using the Common Astronomy Software Applications package (CASA v4.4;  McMullin et al. 2007) by the ALMA observatory. 
The standard calibration  included offline Water Vapor Radiometer (WVR) calibration, system temperature correction, bandpass, phase and amplitude calibrations. 
The observations from all three nights were concatenated and processed together to increase the signal to
noise and \textit{uv}-coverage. 
The visibility data were also edited, calibrated, and imaged using CASA v4.4.
We used the CLEAN algorithm to image the data, and using a robust parameter equal to zero, a 
briggs weighting was performed to adjust balance between resolution and sensitivity. For the continuum, 
we obtained a rms of 0.07 mJy beam$^{-1}$ and a synthesized beam of 0.25$''$ $\times$ 0.17$''$ with P.A. = $-$85 deg. 
For the  line data,  the rms is 12.5 mJy beam$^{-1}$
 for $^{12}$CO, 16.0 mJy beam$^{-1}$ for $^{13}$CO and 13.9 mJy
beam$^{-1}$ for C$^{18}$O with slightly larger beam sizes, $\sim$0.37$''$ $\times$ 0.28$''$.

\begin{figure*}
\includegraphics[width=15cm, trim = 0mm 00mm 0mm 0mm, clip]{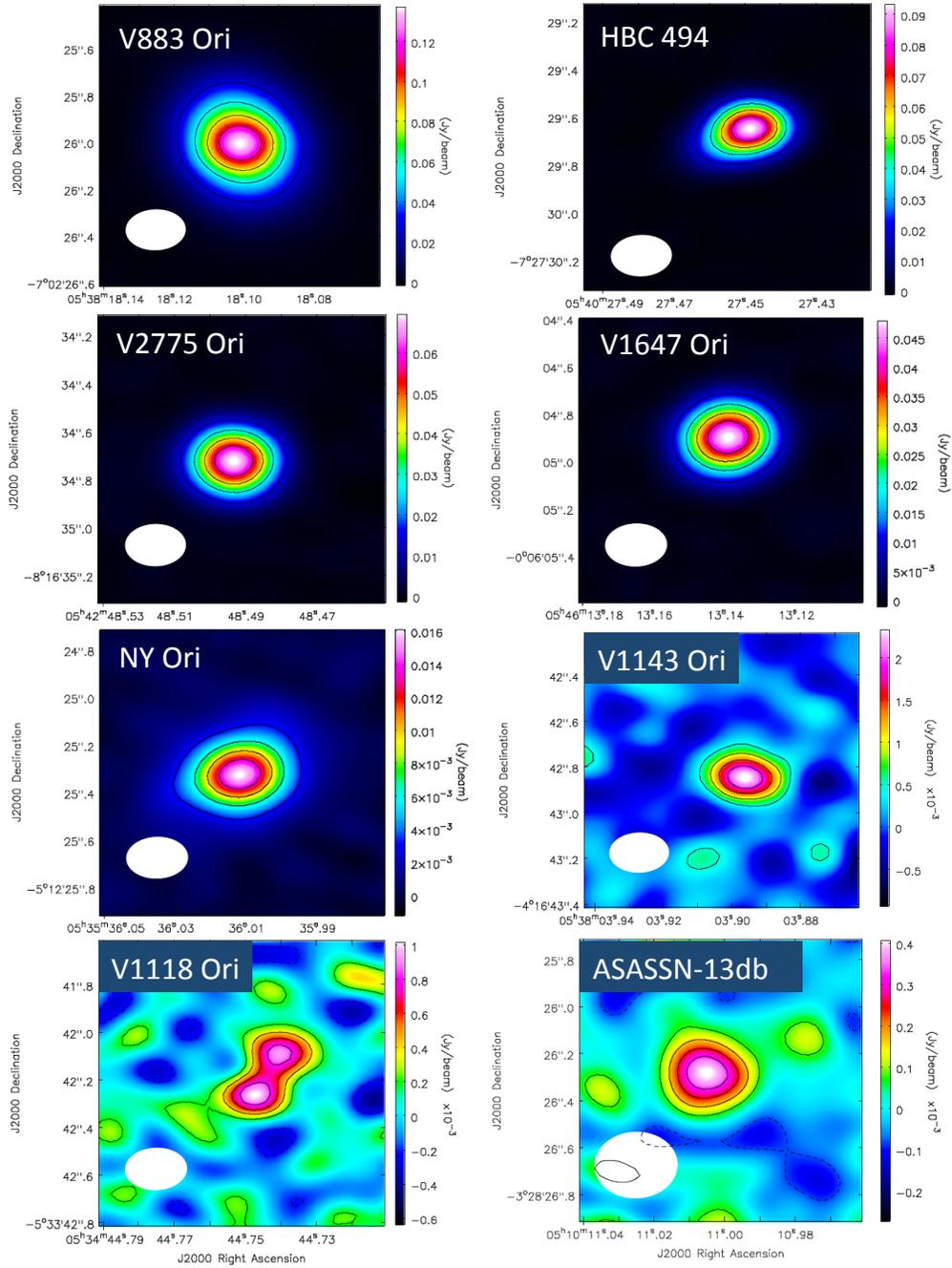}
\caption{ \small 
Continuum 230 GHz/1.3 mm images of our eight targets.   All targets are clearly detected, and the five brightest ones (V883 Ori, HBC~494, V2775 Ori, V1647 Ori and NY Ori) are spatially resolved. 
V1118 Ori is resolved into two millimeters sources, indicating the presence of a disc around each one of the stellar components in the binary system. 
The synthetic beam is shown in the bottom left of each panel.  
}
\label{fig:contimuum}
\end{figure*}

\section{Results}

\subsection{Continuum}\label{continuum}

The continuum images of our targets are shown  in Figure~1. All the objects are clearly detected. V883 Ori, V2775 Ori, HBC~494,  V1647 Ori, and NY Ori have very high signal to noise ratios ($>$ 200) and are clearly resolved.  
We use the IMFIT routine within CASA to fit two dimensional Gaussians to the continuum data and derive both continuum fluxes and disc sizes (deconvolved from the beam) for the resolved sources.
From the ratios of the minor to the major axes, we also derive disc inclinations (see Table 1).  
V1118 Ori, the only close binary  (r $\sim$ 0.18$''$) in our sample, is resolved as two distinct millimeter sources, indicating the presence of small, individually unresolved discs around each one of stellar components. The southern component is slightly brighter in H$\alpha$ ($\Delta$m $\sim$ 0.4 mag; Reipurth et al. 2007) and at 1.3 mm, but it is unclear which component is the source of the outbursts. 
The discs around V1143 Ori and ASASSN--13db are also small and consistent with point-sources (i.e., their sizes cannot be measured but they are likely to be smaller than the beam, $\sim$40 au in radius ). 
Our fluxes for  NY Ori and V1118 Ori agree well with those presented by Liu et al. (2016) based on shallower and lower-resolution observations with the Submillimeter Array (28$\pm$2.2 mJy and 2.0$\pm$0.7 mJy, respectively).  They also observed V1143 Ori, but only report a 3-$\sigma$ upper limit of 1.9 mJy, which is close to the actual value of our 20-$\sigma$ detection. 
Similarly, the flux obtained for V883 Ori agrees (within the 10$\%$ calibration uncertainty) with the flux obtained from ALMA long-baseline observations at 12 au resolution (Cieza et al. 2016), suggesting that our disc fluxes at 80 au resolution are not affected by significant envelope contamination. 

We find that the continuum disc fluxes span over 3 orders of magnitudes from V883~Ori at 350 mJy to ASASSN--13db with a 4-$\sigma$ detection at 0.3 mJy. With the exception of 
ASASSN--13db, the uncertainty of all the fluxes  are dominated by the absolute calibration uncertainty, estimated at the 7$\%$ level for band-6 observations (ALMA Technical Handbook for Cycle 3).
The sizes of the continuum emission  (deconvolved from the beam) for the 5 brightest targets are remarkably small given the disc luminosities and imply disc radii in the $\sim$30--60 AU range. 
Our measurements indicate that the three FUor objects V883~Ori, V2775~Ori, and HBC~494  are significantly brighter than the EXor targets ($\gtrsim$ 100 mJy vs $\lesssim$ 30 mJy).
Interestingly,  V1647 Ori which is considered a borderline case between FUor and EXor objects, has a flux ($\sim$80 mJy) that is intermediate between the two classes of objects. 

\begin{table*}
 \centering
 \begin{minipage}{140mm}
  \caption{Detected discs  (sorted by declining disc mass)}
  \label{table:detections}
  \begin{tabular}{@{}ccccccccc@{}}
  \hline
     & V883 Ori  &  HBC 494  & V2775 Ori & V1647 Ori  & NY Ori &  V1143 Ori & V1118 Ori & ASASSN-13db \\
 \hline
Object Type  & FUor & FUor & FUor & FUor/EXor & EXor & EXor & EXor & EXor  \\
\smallskip
Spectral Type   & ......  & ...... & ...... & ......  & G6-K2 & ......  & M1 & M5 \\ 
\smallskip
L$_{Bol}$ [L$_{\odot}$] & 400 & 300 & 22-28 & 34-44 & ...... & ...... &  7.25 &  0.2 \\
\smallskip
Companions & ...... & ......  & Y? (11$''$) &  ...... &  N &...... & Y (0.18$''$)  &  ......  \\
\smallskip
F$_{1.3mm}$ [mJy]    & 353$\pm$35 &  113$\pm$11 & 106$\pm$10 & 78$\pm$8 &  32$\pm$3 & 2.3$\pm$0.2  &  1.3$\pm$0.1 &  0.3$\pm$0.07 \\
\smallskip
Major axis [$mas$] &  307$\pm$3 & 143$\pm$3 &  151$\pm$1  & 174$\pm$2  & 276$\pm$4 &  unresolved  & unresolved & unresolved \\
\smallskip
Minor axis [$mas$] & 247$\pm$4 & 47$\pm$8 &  146$\pm$2 & 154$\pm$2  &  194$\pm$6  &   unresolved  & unresolved & unresolved \\    
\smallskip
Position Angle [deg] & 32$\pm$3 &  120$\pm$2 & 48$\pm$26  &  139$\pm5$&  113$\pm$2 &  ...... & ......  &......  \\
\smallskip
disc radius [au]      &    64    & 30 &    31  &  36  &  57  & $<$ 40 & $<$ 40 & $<$ 40   \\
\smallskip
Inclination [deg] & 37 & 70 & 15& 28 & 45 & ...... &...... &...... \\
\smallskip 
M$_{DUST}$ [M$_{\oplus}$]     & 1800 & 570 & 530  & 390 & 160 & 12 & 6.5 &  1.5  \\
\smallskip
M$_{disc}$  [M$_{JUP}$]   & 570 & 180  &  170  & 120 &  50 &  3.8 &  2.0  &  0.5 \\     
\hline
\end{tabular}
The data for the four top rows (Object Type, Spectral Type, and L$_{Bol}$, and Companions) are taken from the review by Audard et al. (2014) for all sources except for ASASSN--13db, the data of which come from  Holoien et al. (2014). 
\end{minipage}
\end{table*}

\subsection{Line data}

The molecular line data for the three FUor  objects in our sample and V1647 Ori, all of which show prominent outflows, are discussed in Cieza et al. (2016),  Zurlo et al. (2017),  Ruiz-Rodriguez  (2017a,b) and Principe et al. (in press). 
While all of the targets show $^{12}$CO emission from the ambient cloud, none of the EXor objects display evidence for outflow activity. 
For completeness, we report that the NY Ori disc is detected as a compact source in $^{13}$CO and C$^{18}$O with fluxes of 1.0  and 0.3 Jy km s$^{-1}$, respectively. 
For V1143 Ori, V1118 Ori,  and ASASSN--13db, no line emission is detected from their discs at the line sensitivity of our survey (3-$\sigma$ $\sim$50 mJy km s$^{-1}$).

\section{Discussion}
\subsection{Dust masses}

Except for the inner regions of very massive systems, protoplanetay discs become optically thin at  millimeter wavelengths.
This implies that most dust grains contribute to the  observed emission and the total flux correlates well with  the total mass of small grains (size $\sim$ $\lambda$).  
Millimeter fluxes can thus be used to estimate the dust masses of protoplanetary discs  using the following formula: 

\begin{equation}
M_{dust}  =  \frac{F_\nu d^2}{\kappa_\nu B_\nu (T_{dust})} 
\end{equation}

where $d$ is the distance to the target,  $T$ is the dust temperature  
and  $\kappa_{\nu}$ is the dust opacity.
Adopting the distance of 414$\pm$7 pc from Menten et al. (2007) to the Orion Nebula Cluster and making  standard assumptions about the disc temperature (T$_{dust}$ = 20 K)  
and dust opacity ($\kappa_{\nu}$ = 10($\nu/$1000 GHz)cm$^2$g$^{-1}$; Beckwith et al.  1990),  Equation~1 becomes:

\begin{equation}
M_{dust}  [M_{\oplus}] =  5.0 \times F_{1.3}  (mJy)
\end{equation}

The derived dust masses are listed in Table 1, but we note that  significant uncertainties exists in the dust opacities, temperatures and distances (our targets are many degrees apart   and distances in the  Orion Molecular Cloud Complex range from 380 to 430 pc; Kounkel et al. 2017)
Also, for the brightest  objects,  the mass fraction of the optically thick material in the inner disc is expected to become 
substantial (Perez et al. 2010), which would result in underestimated dust masses. On the other hand, the disc temperatures might be higher than the  standard 20 K assumed due to viscous heating and the particularly bright central sources, resulting in overestimated dust masses. 
It is not obvious which of these two effects dominates in each object, but they would tend to compensate each other to some extent.
For completeness, we also calculate the disc masses adopting a  standard gas-to-dust mass ratio of 100 (see M$_{disc}$ row in Table ~1) and note that the three FUor objects have nominal disc masses in the $\sim$0.2--0.6~M$_{\odot}$ range. 

\subsection{Radiative transfer modeling}\label{MCMC}

We construct radiative transfer models using the RADMC-3D code (Dullemond et al. 2012), to derive the physical properties of the spatially resolved discs. 
Circumstellar discs around T Tauri and Herbig A/B stars are typically modeled as passive discs heated only by the stellar photosphere; however, in the case of FUor objects, the accretion luminosity is expected to overwhelm emission from the stellar
photosphere. For instance, while the bolometric luminosity of V883 Ori is $\sim$400 L$_{\odot}$, its central object is a 1.3 M$_{\odot}$ star with an expected photospheric luminosity of only $\sim$6 L$_{\odot}$  (Cieza et al. 2016). 
Therefore, we model the central heating source as a 10,000 K blackbody (an A0-type star to be precise)  with a  luminosity equal to the bolometric luminosity of the system to approximate the accretion-shock luminosity.
Viscous heating is ignored, but we note that it might become important in the inner regions of the disc with the highest accretion rates. 
Since we are interested in basic parameters such as the mass, radius,  and surface density profiles, and the comparison to non-outbursting discs, we adopt the parameterization presented by Andrews et al. 2009.
This parameterization is motivated by  accretion theory  (Lynden-Bell $\&$ Pringle 1974; Hartmann et al. 1998) where the surface density, $\Sigma$, is characterized by a power-law with an index $-\gamma$ in the inner disc and an exponential taper at large radii:

\begin{equation}
\Sigma = \Sigma_c  \left( \frac{R}{R_c} \right)^{-\gamma} exp \Bigg[ -   \left( \frac{R}{R_c} \right)^{2-\gamma}  \Bigg]
\end{equation}

\noindent where  R$_c$ is the \emph{characteristic} radius of the disc (as opposed to a sharp outer radius). The scale height as a function of radius is given by:

\begin{equation}
h = h_c  \left( \frac{R}{R_c} \right)^\Psi
\end{equation}

\noindent where h$_c$ is the scale height at a characteristic radius R$_c$, and $\Psi$ defines the degree of  flaring in the disc.
The discs can therefore be described by 5 free parameters,  R$_c$, $\gamma$,  $\Psi$, h$_c$, and $\Sigma_c$.
We integrate Equation 3, and calculate the disc mass as: 

\begin{equation}
M_d =   \frac{2 \pi R_c^2  \Sigma_c   }{2 - \gamma} 
\end{equation}

We adopt dust grains with a standard power-law distribution of grain sizes $a$, given by $n(a)
\propto a^{-3.5}$ and extending from $a_{\rm min} = 0.1\,\mu$m to
$a_{\rm max} = 3\,$mm.  For the dust optical properties, we use a
mass-weighted mean opacity of amorphous carbon grains from (Li \&
Greenberg 1997) and astrosilicate grains (Draine 2003).  Both species
were combined in a mix using Bruggeman's rules. Opacities of the mix
were computed using the 'Mie Theory' code written by Bohren \& Huffman
(1983). The absorption opacity at 1.3~mm is $\kappa_{\rm abs} =
2.2\,{\rm cm}^2\,{\rm g}^{-1}$.

The model parameters $\{M_{\rm disc}, \gamma, R_{\rm c}, h_{\rm c},
\psi\}$ were constrained using a Bayesian approach. In addition to the
disc structure parameters, we also allow the centroid shift ($\delta
x, \delta y$), the inclination angle $i$, and the PA of the model to
vary. 
Since most of the spatially resolved discs correspond to embedded objects,
the information about the stellar spectrum is not well known. 
 Hence, we also explore $R_{\rm
  star}$ values to improve the fit, assuming an effective temperature
of 10000K to account for the stellar photosphere and the accretion
luminosity.

The posterior distribution for each parameter was recovered using a
Goodman \& Weare's Affine Invariant MCMC ensemble sampler
(Foreman-Mackay et al. 2013). We used the publicly available {\sc
  python} module {\em emcee} to sample the parameter space and
maximise the likelihood function. The likelihood function is
proportional to $\exp[-\chi^2/2]$, where $\chi^2$ is the sum over the
squared difference of the model and measured visibilities, divided by
their variance. Additionally, we define a parameter $f_{\rm sigma}$ to
account for the uncertainty on the weights of the observed
visibilities. Varying $f_{\rm sigma}$ aims to produce a reduced
$\chi^2=1$, allowing for more meaningful uncertainties to be drawn
from the posteriors of each parameter. Visibilities weights are
therefore divided by $f_{\rm sigma}^2$.

Maximising the likelihood function is equivalent to minimising the
negative of its logarithm. Therefore, we aim to minimise the following function:
\begin{equation}
  -\log{P(\Theta)} = \frac{1}{2}\frac{\chi^2}{f_{\rm sigma}^2} + N_{\rm vis} \log{f_{\rm sigma}},
\end{equation}

\noindent where $P(\Theta)$ is the likelihood function, $\Theta$ are
the free parameters, and $N_{\rm vis}$ is the number of
visibilities.
We used the results from the two dimensional Gaussians fits (see Table 1) as initial guesses for the 
disc parameter of the resolved objects and explored the parameter space around them. In particular, 
our priors for the free parameters come from assuming
uniform distributions in the following intervals. 

\begin{eqnarray*}
  M_{\rm disc} &\in& [0.001, 1.0]\, M_\odot,\\
  \gamma &\in& [0.0, 2.0]\, M_\odot,\\
  R_{\rm star} &\in& [1.0, 8]\, R_\odot,\\
  R_{\rm c} &\in& [10.0, 100.0]\, {\rm au},\\
  h_{\rm c} &\in& [1.0, 10.0]\, {\rm au},\\
  \psi &\in& [0.0, 2.0]\, {\rm au},\\
  \delta x &\in& [-0.1, 0.1]'', \\ 
  \delta y &\in& [-0.1, 0.1]'', \\
  i &\in& [10, 80]^\circ, \\ 
  {\rm PA} &\in& [10, 180]^\circ,\\
  f_{\rm sigma} &\in& [0.0, 10.].
\end{eqnarray*}

\begin{figure*}
  \includegraphics[width=1.1\textwidth, trim = 0mm 60mm 0mm 0mm, clip]{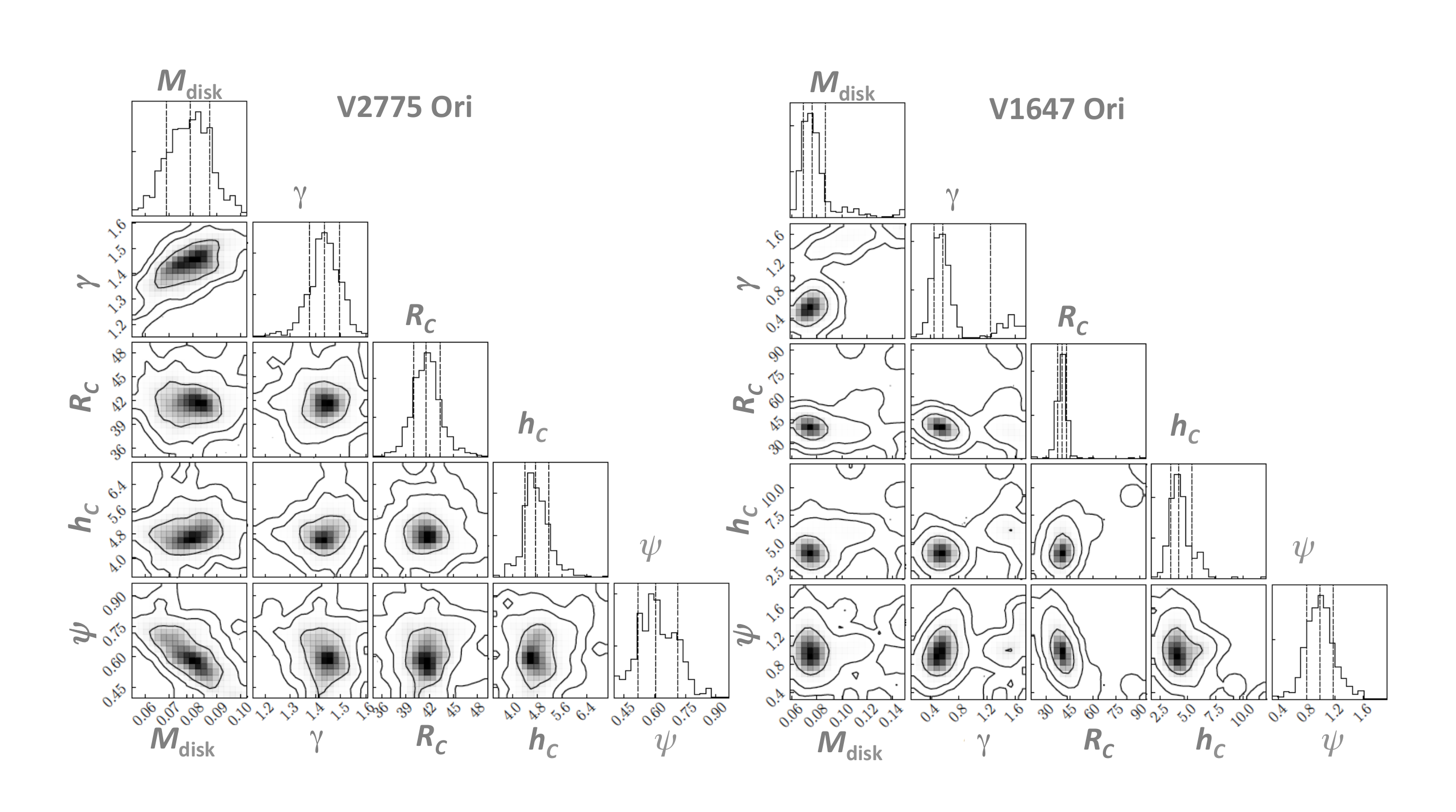}
 \includegraphics[width=0.55\textwidth, trim = 10mm 10mm 80mm 10mm, clip]{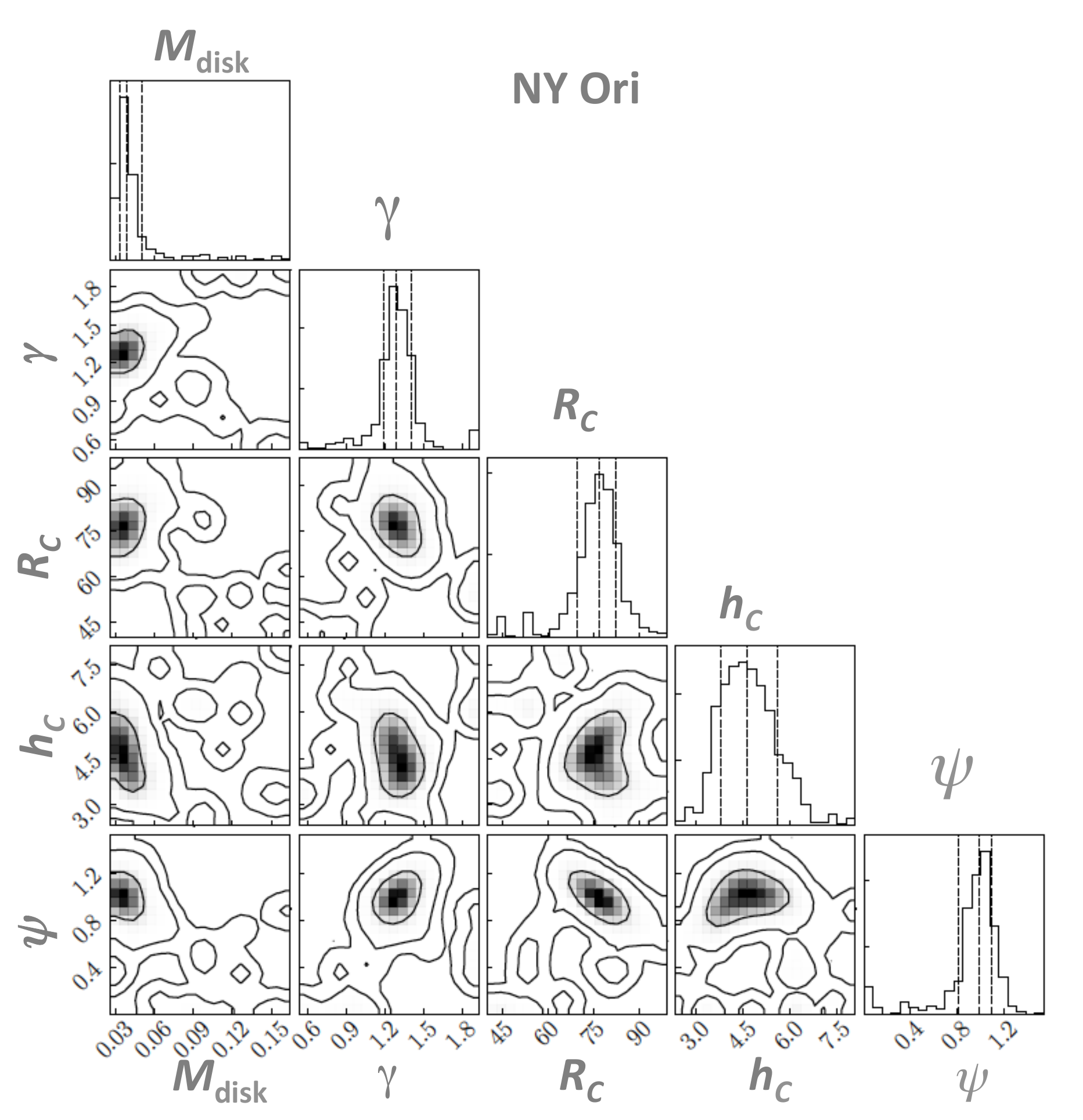}
  \caption{Posterior distributions of each disc parameter, including
    their marginalised distributions for  V2775 Ori, V1647 Ori, and  NY Ori. The vertical dashed lines
    represent the 16th, 50th and 84th percentiles. Contours correspond
    to 68 per cent, 95 per cent and 99.7 per cent confidence regions.
    These plots were produced using the {\sc python} module {\rm
      corner} (Foreman-Mackey et al. 2014).}
  \label{fig:MCMC1}
\end{figure*}

\begin{figure*}
  \includegraphics[width=1.1\textwidth, trim = 0mm 60mm 0mm 0mm, clip ]{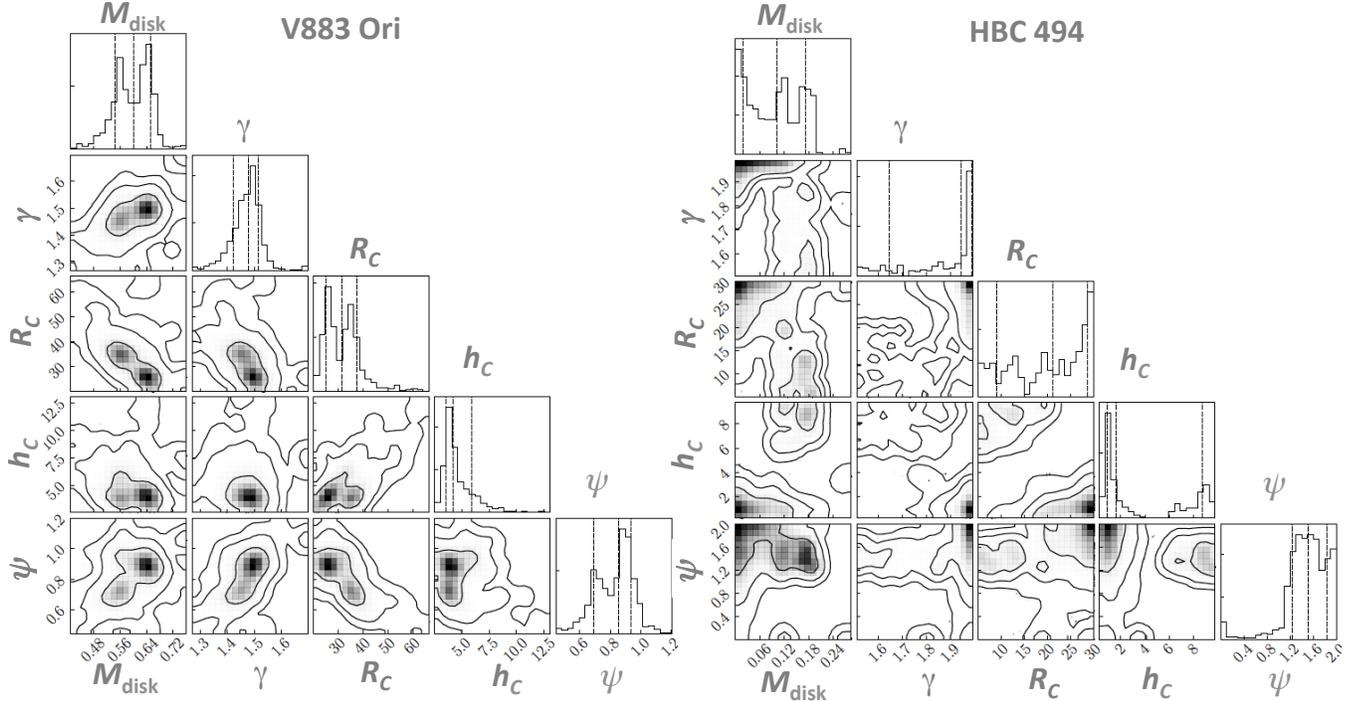}
  \caption{
  Same as Figure 2, but for V883 Ori and HBC~494.}
  \label{fig:MCMC2}
\end{figure*}

\begin{figure*}
  \centering\includegraphics[width=.8\textwidth]{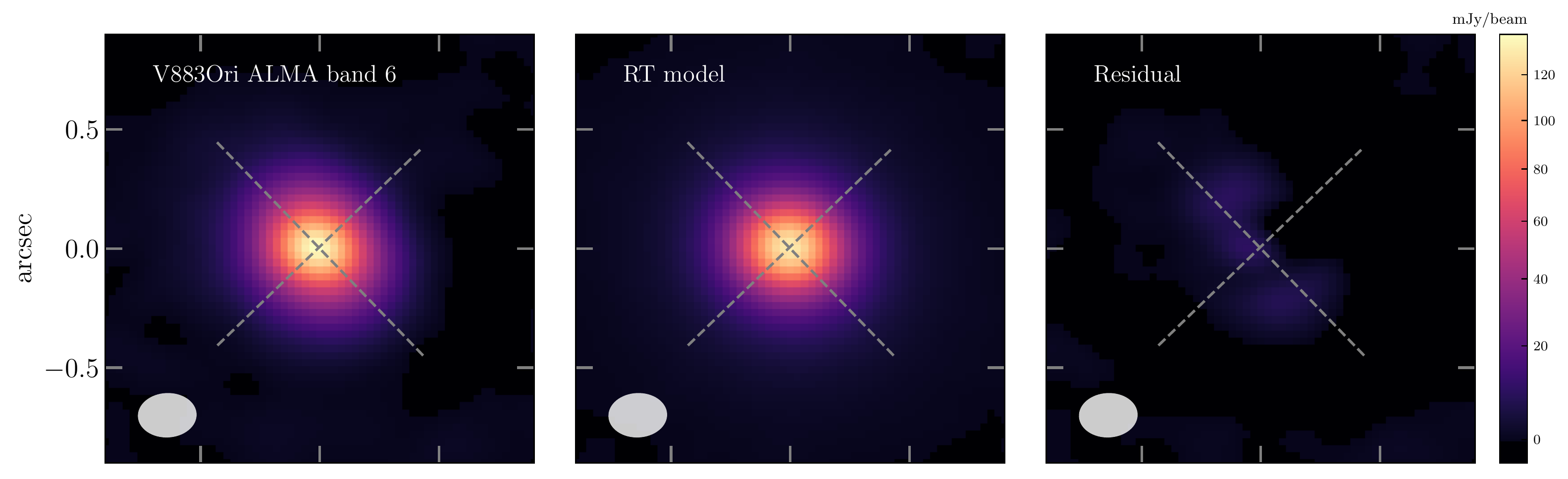}
  \centering\includegraphics[width=.8\textwidth]{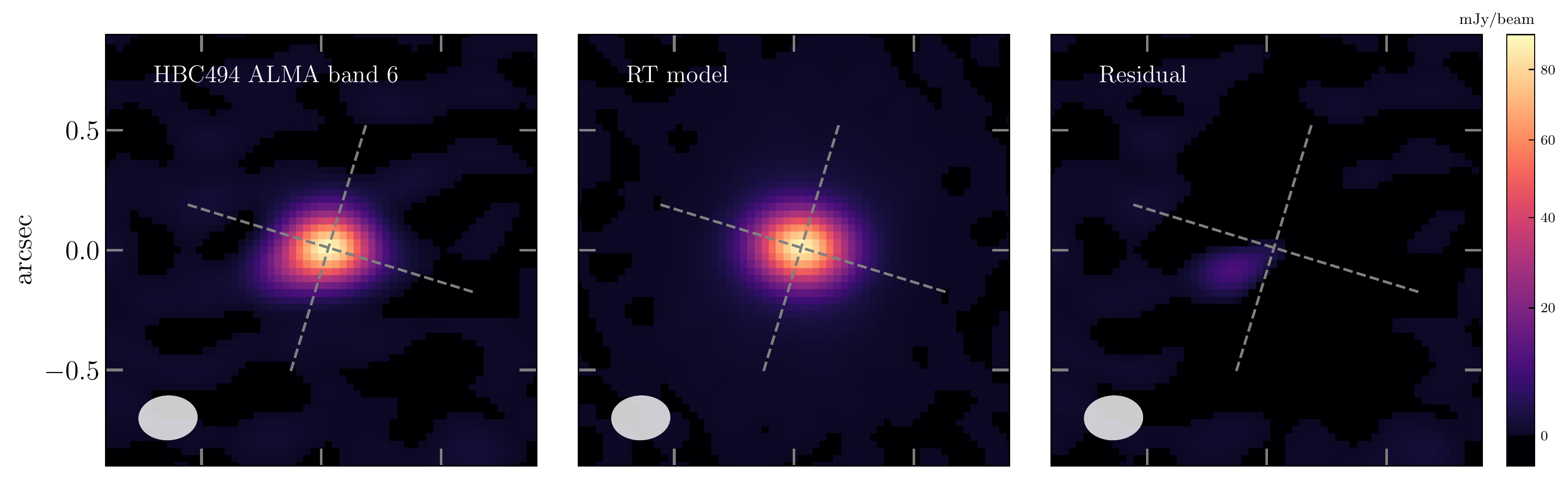}
  \centering\includegraphics[width=.8\textwidth]{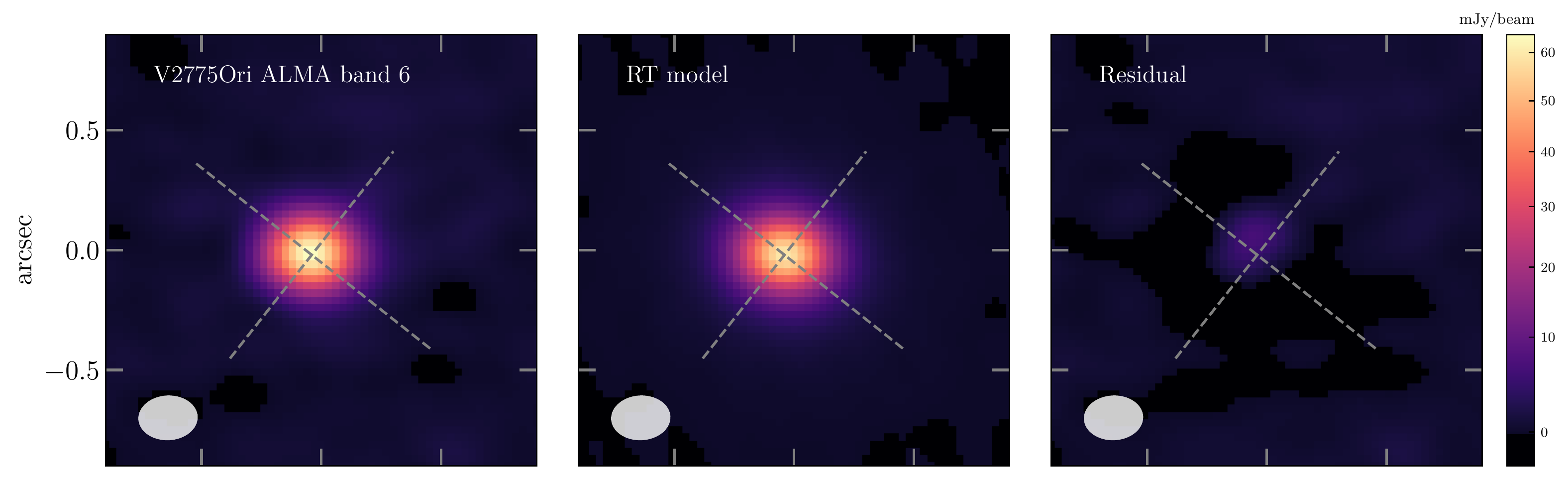}
  \centering\includegraphics[width=.8\textwidth]{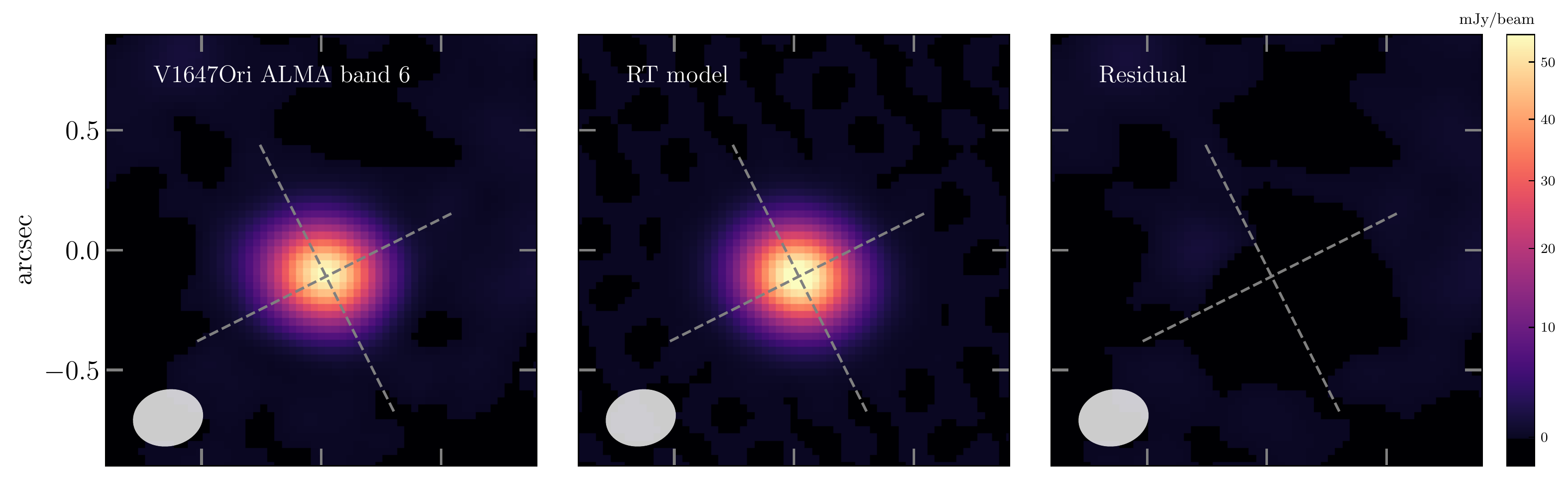}
  \centering\includegraphics[width=.8\textwidth]{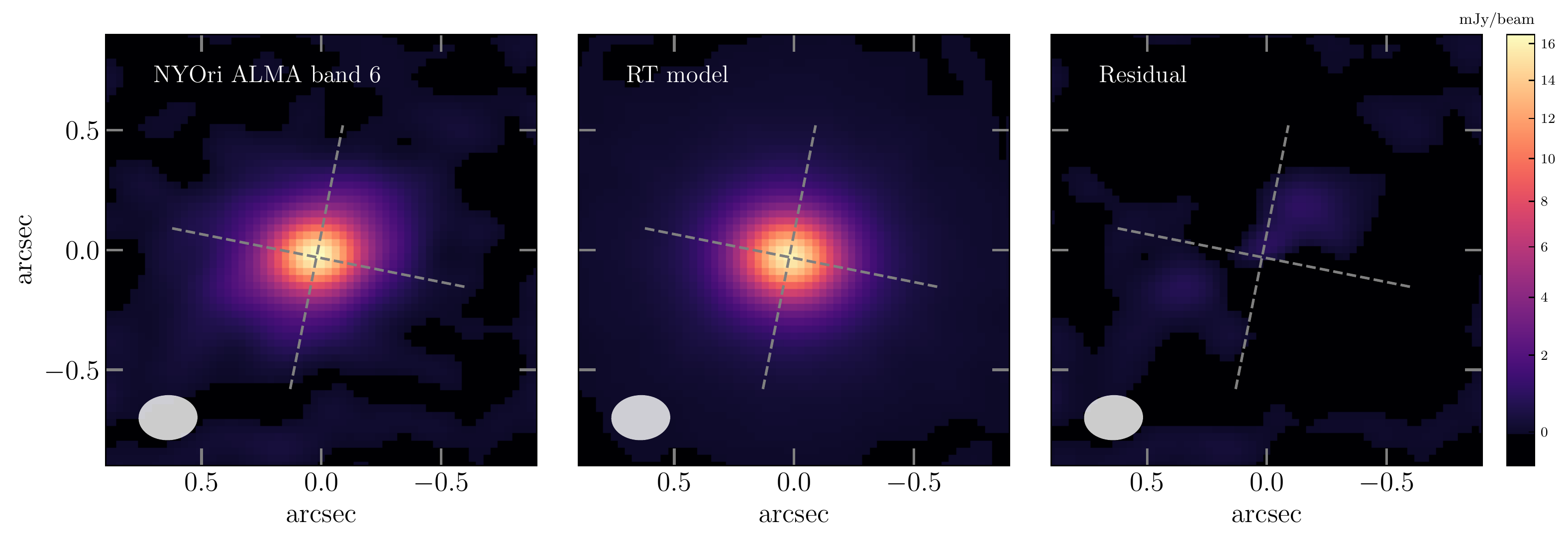}
  \caption{The ALMA 1.3 mm continuum data (left), synthetic best-fitting models (middle) and 
  the residuals (right) for the 5 resolved discs in our sample.}
  \label{fig:bestfit}
\end{figure*}

The best-fitting parameters and their uncertainties were obtained
after running 1000 iterations ($\sim10$ times the autocorrelation
time) with approximately 100 walkers. The model visibilities are
obtained by taking the Fast Fourier Transform of model images and
interpolating to the same \textit{uv} points as the observations (Marino et
al. in prep). Each measurement set (5 brightest discs) is fit
separately.
%
Posterior distribution of $\{M_{\rm disc}, \gamma, R_{\rm c}, h_{\rm
 c}, \psi\}$ are presented in Figures~\ref{fig:MCMC1} and~\ref{fig:MCMC2}, while the best-fit models are shown in Figure~\ref{fig:bestfit}.

\begin{table*}
\caption[]{\em{Disc parameters from Radiative Transfer Modeling}}
\begin{tabular}{lcrccccc}
\hline \noalign {\smallskip}
Name & R$_{c}$ & M$_{disc}$  & $\gamma$   & h$_{c}$ & $\Psi$  \\ 
    & (au) & (M$_{JUP}$)&  &  (au) &  \\
\hline \noalign {\smallskip}
V883 Ori     &   31$^{+6}_{-6}$   & 600$^{+50}_{-60}$        & 1.48$^{+0.04}_{-0.06}$        & 4.2$^{+1.8}_{-0.6}$           & 0.86$^{+0.08}_{-0.16}$  \\
HBC 494     &  21$^{+7}_{-12}$ & 100$^{+70}_{-80}$        & 1.94$^{+0.04}_{-0.30}$         & 1.6$^{+7.1}_{-0.8}$           & 1.52$^{+0.32}_{-0.28}$    \\
V2775 Ori   &  42$^{+2}_{-2}$   &   80$^{+10}_{-10}$         & 1.44$^{+0.06}_{-0.06}$         & 4.7$^{+0.4}_{-0.3}$          & 0.60$^{+0.11}_{-0.08}$   \\
V1647 Ori   &  40$^{+3}_{-3}$   &   80$^{+10}_{-10}$         &  0.57$^{+0.68}_{-0.13}$        & 4.2$^{+1.1}_{-0.7}$          & 0.99$^{+0.18}_{-0.19}$  \\
NY Ori         &  76$^{+5}_{-7}$   &   40$^{+10}_{-10}$         &  1.29$^{+0.11}_{-0.10}$        & 4.6$^{+1.0}_{-0.8}$          & 0.98$^{+0.12}_{-0.17}$    \\
\hline \noalign {\smallskip}
\end{tabular}\label{table:param}
\end{table*}

\subsubsection{Radiative transfer results}\label{RTresults}

\begin{figure*}
  \includegraphics[trim = 200mm 0mm 0mm 0mm, clip, width=0.4\textwidth]{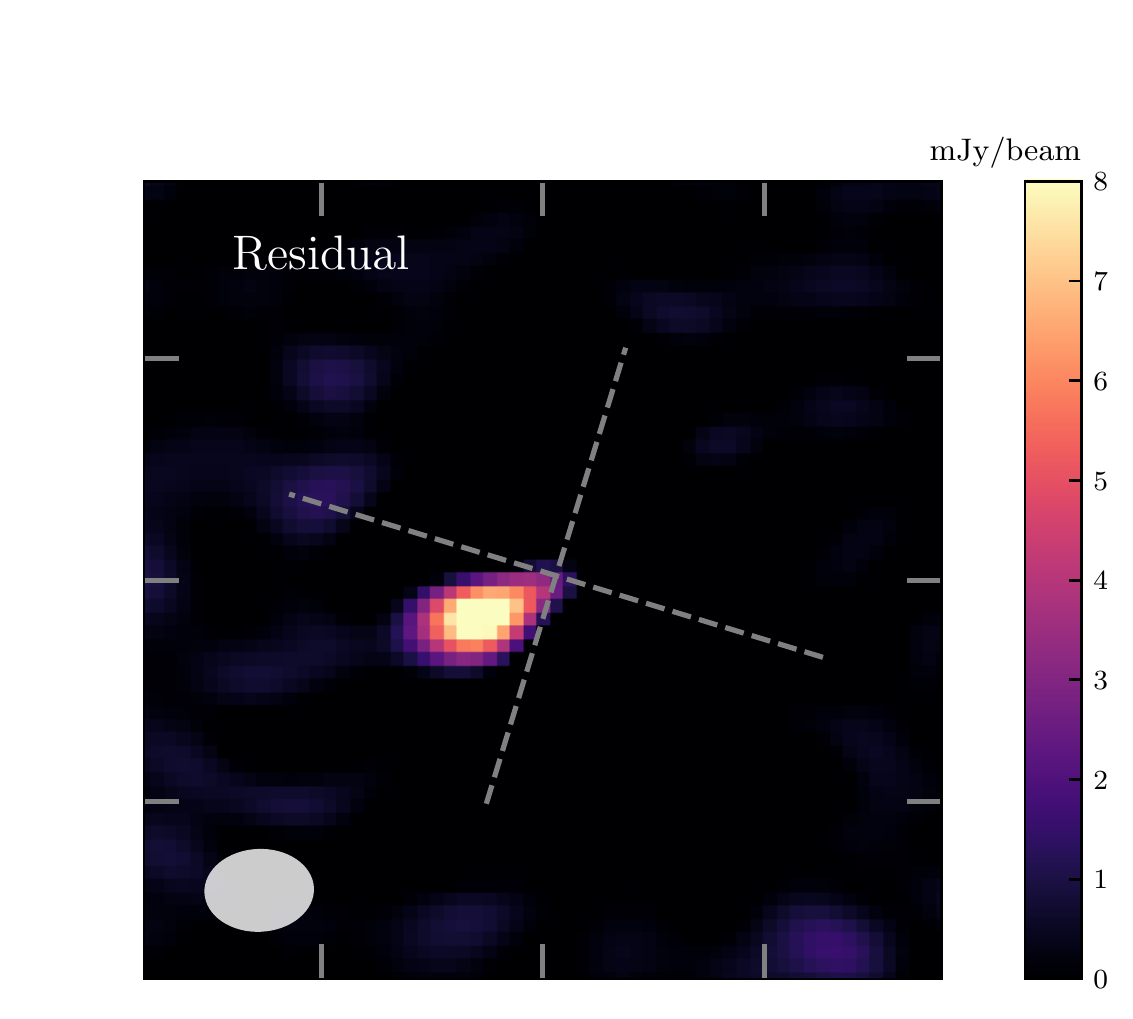}
  \caption{The ALMA data of HBC~494 minus our best-fit model.  The flux of this residual is $\sim$8 mJy or $\sim$7$\%$ of the total flux of the system. This emission might correspond to  a compact structure (non point-symmetric) in the disc, or to a second, fainter disc around a stellar companion.}
 \label{fig:hbc494}
\end{figure*}

We find that, at the observed resolution of 0.2$''$, V2775 Ori, V1647 Ori, and NY Ori are well described by our parametric models: the posteriori distributions of the disc parameters are relatively narrow and single-peaked (see Figure~\ref{fig:MCMC1}). 
However, V883~Ori and HBC~494 have disc parameters that are less well-defined. In the case of V883 Ori, the posteriori distributions for $M_{disc}$, $R_c$, and $\Psi$ are double-peaked (see Figure~\ref{fig:MCMC1}, left panel). We speculate this is due to the particular structure of V883 Ori, showing a very bright core, 0.2$''$ in diameter, surrounded by a much more tenuous outer disc.  The boundary between these two regions has been interpreted as the water snow-line, resulting in a discontinuity in the disc properties (Cieza et al. 2016). The water snow-line is typically at $\sim$3-5 au for solar-mass stars and would unidentifiable at the 80 au resolution of our observations,  but given the extreme accretion rate of  V883 Ori, it has been displaced to a radius of $\sim$40 au.  
Detailed modeling of the V883 Ori disc at $\sim$0.03$''$ (15 au)  resolution will be presented in a follow-up paper. 
Similarly,  in the case of HBC~494,  the posteriori distributions for M$_{disc}$ and $R_c$ are wider than for the other sources (see Figure~\ref{fig:MCMC2}, right panel). 
This is likely due to the fact that the disc is not radially symmetric, preventing the MCMC  exploration of parameters to converge to a narrower range of values. 
The disc asymmetry in HBC~494 can be seen in Figures \ref{fig:contimuum} and \ref{fig:bestfit} as an elongation of the disc on the south-east direction. 
We speculate that this asymmetry  might indicate the presence of two barely resolved discs (e.g., the system might be a binary as  V1118 Ori and FU Ori itself; Hales et al. 2015; Liu et al. 2017) or  a non-axisymmetric disc structure.
Figure~\ref{fig:hbc494} shows a $\sim$8 mJy residual in our ALMA image after subtracting the best-fit model, but in different scaling than Figure~\ref{fig:bestfit}.  
This residual has the same intensity level ($\sim$7$\%$ of the total flux) than in other sources, but is compact an not radially symmetric. 
A particularly interesting possibility for this residual would be the presence of a gravitational instability clump, which could be bright enough to be detected with ALMA (Zhu et al. 2012).
ALMA images of the  HBC~494 system at $\sim$0.03$''$ (15 au) resolution will be presented in a follow-up paper.  

We note that the disc masses derived from the radiative transfer modeling agree with the values calculated from Equation~2 (see Table~1) 
to within 5 and 20$\%$ for the brightest and faintest objects we modeled (V883 Ori and NY Ori, respectively), but are up to $\sim$50$\%$ lower for HBC 494 and V2775 Ori. 
Similarly, some of the $R_{c}$ values agree well (e.g., to within 10$\%$ in the case of  V1647 Ori)  with the discs sizes derived from the two dimensional Gaussians fitting  (also in Table 1), but can be up to 50$\%$ smaller for some objects like V883 Ori. 
These differences underline the importance of using similar procedures when comparing the properties of different discs.

\subsubsection{Comparison between FUor, EXor, and T Tauri discs}

\begin{figure*}
\includegraphics[angle=0, width=15cm, trim = 10mm 10mm 10mm 10mm, clip]{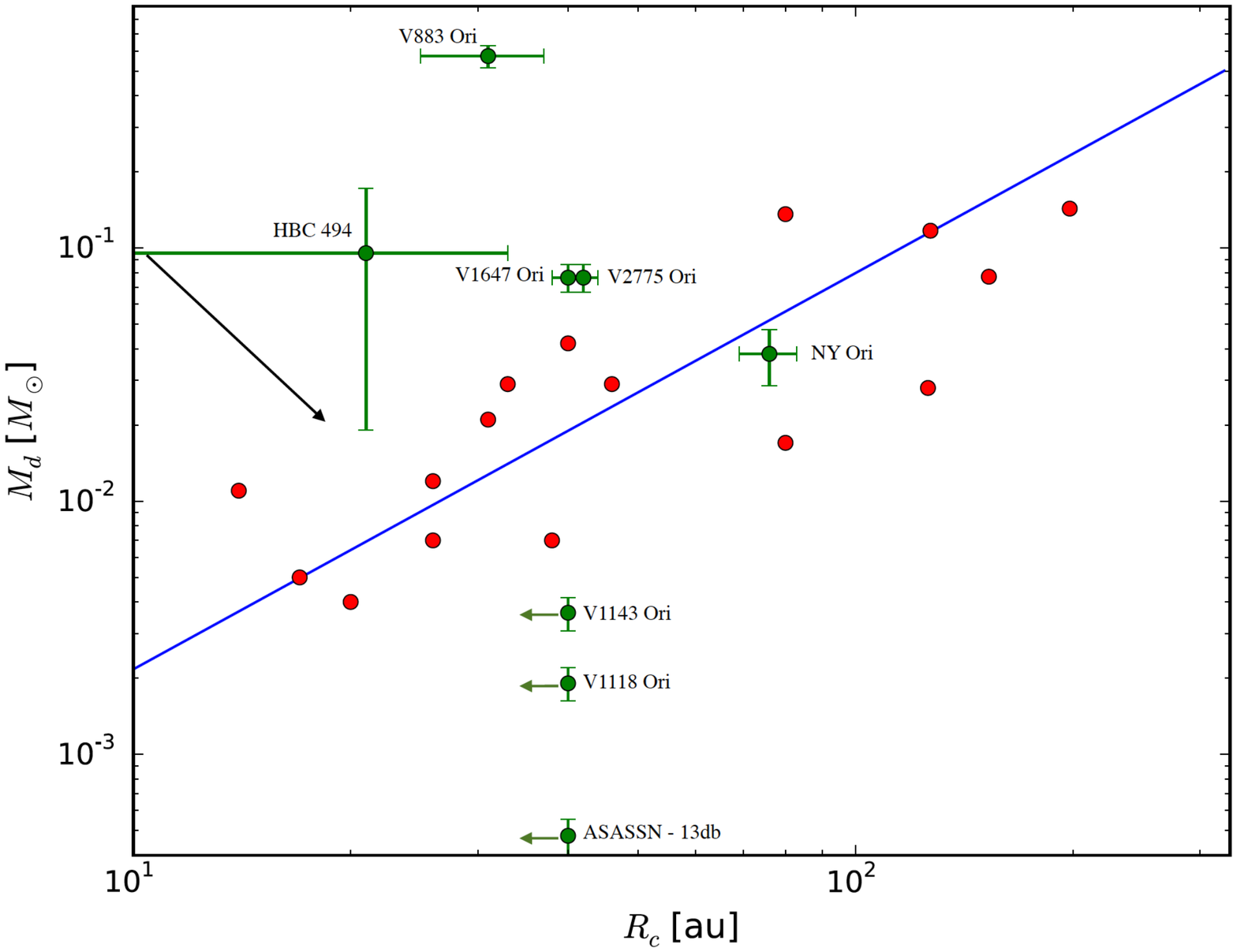}
\caption{ \small 
Disc masses as a function of R$_c$ (and upper size limits) for our 8  targets (green points) compared to the sample of T Tauri stars  (red points) studied by Andrews et al. (2010). 
We find that the FUor objects (V883 Ori, HBC 494, and V2775 Ori) have much larger masses than the EXor sources (NY Ori, V1143 Ori, V1118 Ori and ASASSN-13db).
The borderline object V1647 has very similar disc properties to V2775 Ori. 
The FUor sources have larger disc masses  for a given disc size than the T Tauri stars.  NY Ori falls in the trend followed by T Tauri stars (blue line). The other EXor objects have no size measurements, but they are unlikely to fall above this  trend, unless they have tiny discs that are less than 0.1-5 au in radius. 
The black arrow shows the general evolution expected from the conservation of angular momentum in a viscous disc,  given by M$_d$ $\propto$ R$_c^{-1/2}$. 
The FUor objects in our sample are embedded Class I stars that might evolve onto the T Tauri locus as they evolve.   
}
\label{fig:mdvsrc}
\end{figure*}

Using the same parameterization and basic assumptions adopted here (opacities and gas to dust mass ratio), Andrews et al. (2010) found that T Tauri stars disc show a correlation between R$_{c}$ disc mass such that M$_d$ $\propto$ R$_c^{1.6}$. 
Figure~\ref{fig:mdvsrc} places our sample of outbursting sources in the M$_d$ vs. R$_c$ plane together with the objects studied by Andrews et al. (2010).
We find that the discs of the FUor sources have very small radii for their disc masses. 
We also find that FUor  in our sample tend to have large values of $\gamma$ compared with other T Tauri stars. 
The 16 T Tauri stars studied by Andrews et al. (2010) have $\gamma$ values ranging from 0.4 to 1.1 with a mean of 0.87$\pm$0.17. 
The $\gamma$ values of our FUor sources are 1.48$^{+0.04}_{-0.06}$, 1.94$^{+0.04}_{-0.30}$,  1.44$^{+0.06}_{-0.06}$ and therefore larger than those of \emph{any} of the T Tauri stars.
This implies that FUor sources tend to have discs that are massive but small and centrally concentrated. 
This structure could be related to their evolutionary status (young  Class I sources) as conservation of angular momentum in a viscous discs dictates that discs should start small and spread out as they accrete onto the star, following M$_d$ $\propto$ R$_c^{-1/2}$.
Dust drifting usually results  in discs that have larger radii as identified via gas tracers compared with their dust radius as traced via (sub)millimeter continuum (Pietu et al. 2005; Isella et al. 2007);   therefore, the trend of large disc sizes with age might be even stronger in the gas than what is seen in the dust. 
It is also possible that the compact and centrally peaked nature of these targets is a property of the FUor sources related to their outburst mechanism. 
Distinguishing both scenarios (intrinsic vs evolutionary differences)  requires larger samples of T Tauri, Class I, and FUor discs for which R$_c$ and M$_{disc}$ values could be calculated in a consistent way.

\subsection{Implications for Outburst Mechanism}

As discussed in \S~1, several different mechanisms  have been proposed to explain the outburst phenomenon in FUor and EXor sources.
Some of these mechanisms, such as the coupling of magnetorotational  and gravitational instabilities (MRI + GI, Armitage et al. 2001, Zhu et al. 2009, Martin et al. 2012) and 
disc fragmentation (Vorobyov $\&$ Basu 2005, Zhu et al. 2012) require  massive discs (M$_{disc}$/M$_{\star}$ $\gtrsim$ 0.1).  
Other outbursts mechanisms, including thermal instability induced by planets (Clarke et al. 1990, Lodato $\&$ Clarke 2004) or external triggers  such as close stellar encounters  (Bonnell $\&$ Bastien 1992) are less dependent on disc mass. 

One of the main results of our small survey is the large range in millimeter fluxes of our outbursting targets and the significant difference in disc masses of FUor and EXor sources. 
%
%
This is an indication that FUor and EXor outburst represent different stages of disc evolution, but also has implications for some potential outburst mechanisms.
In particular, the disc masses of the three FUor objects (and perhaps of V1647 as well) are of the order of 0.1--0.6 M$_{\odot}$ and could be susceptible to Gravitational Instability. 
However, V883~Ori, the brightest disc in our sample,  was observed at 30 mas (12 au) resolution at 230 GHz and shows no signs of disc fragmentation (it has an axisymmetric feature at  40 au, consistent with the water snowline, Cieza et al. 2016). 
Similarly,  HL Tau, with an estimated   M$_{disc}$/M$_{\star}$ $\sim$ 0.1   (Tamayo et al. 2015),   shows no signs of instability (spiral arms or clumps) even when observed at 5 AU resolution (ALMA Partnership et al. 2015). 
The disc around Elias 2-27 does show  spiral arms similar to those predicted by gravitational instability; however, the object is not outbursting and the calculation of its Toomre Q instability parameter seems to indicate that the disc is \emph{not} gravitationally unstable (Perez et al. 2016). 
Gravitational instability has recently been imaged in a Class~0 object forming a multiple star system (Tobin et al. 2016), 
but has never been seen in a Class I disc or in a context that could explain the FUor phenomenon. 
As discussed in \S~\ref{MCMC}, HBC~494 is a candidate for disc fragmentation through gravitational instability (see Sec~\ref{MCMC}) and higher resolution images are needed to explore this possibility.  
Overall, models that combine MRI and GI without fragmentation  (e.g., Zhu et al. 2009) seem the most consistent with the observational properties of FUor sources.

For the EXor objects, we find disc masses similar to those of normal T Tauri stars. In these cases, GI is unlikely to play an important role in the outbursts, and our results suggest that other mechanisms, such as dynamical perturbations,  must be responsible for the EXor phenomenon. 
From our millimeter imaging, only V1118~Ori, which was already known to be a binary system, was clearly resolved into two components.  As discussed in Sec~\ref{MCMC},  HBC~494 might also be a binary, pending confirmation from higher resolution observations. 
However, studies trying to connect disc outburst with multiplicity have so far been inconclusive (e.g.,  Millan-Gabet et al. 2006; Green et al. 2016) and there is no clear evidence that  the multiplicity statistics from outbursting sources are any different from those of non-outbursting young stellar objects. 
Everything considered (low-disc mass and lack of clear connection with multiplicity), it seems likely that the EXor phenomenon is connected to instabilities of the inner disc 
(e.g.,  D$'$Angelo $\&$ Spruit 2012) and/or planet-disc interactions (Clarke et al. 1990, Lodato $\&$ Clarke 2004). 
 
An important upstanding issue is whether most or all young stellar objects go through  FUor and/or EXor  episodes.  Solving the ``luminosity problem"  of the general population of protostars (Kenyon et al. 1990, Evans et al. 2009)  through episodic accretion requires that most of them go through these outbursts at some point of their evolution to build enough mass. However, this general requirement is degenerate between the accretion rate, the duty cycle, and the duration of  the outbursts. 
The review by Audard et al. (2014) lists a larger number of FU Or objects than EXors, but the former are more conspicuous and easier to detect that the later group and their census should be more complete for a given volume.   Hillenbrand et al. (2015) argues that current observational constraints are consistent with an FUor outburst rate of 10$^{-5}$ year$^{-1}$ star$^{-1}$ independently of whether all or some young stellar objects go through this phase. They estimate that constraining the occurrence of accretion outburst down to an accretion rate sensitivity of 10$^{-4}$ M$_{\odot}$ yr$^{-1}$ would required monitoring 10$^{5}$ young stars for 1 year or 10$^{4}$ stars for 10 years.  
Similarly, constraining the occurrence of accretion outburst down to a sensitivity of 10$^{-6}$ M$_{\odot}$ yr$^{-1}$ (which would include EXors) requires monitoring 10$^{5}$ young stars for 10 years.  These requirements are within the reach of the Large Synoptic Survey Telescope. 

\section{Summary and Conclusions}

As part of a series of papers investigating the circumstellar properties of outbursting stars in Orion, we present ALMA dust continuum images in band-6 (230 GHz) at 0.2$''$  (80 au) resolution of three FUor, four EXor and one borderline object. We detect all of our targets and resolve 5 of them, for which we perform radiative transfer modeling. From the analysis of our data, we derive the following conclusions: \\

%

\noindent 1) The millimeter wavelength fluxes (and the derived disc masses) of these outbursting objects span over 3 orders of magnitude. Even in our small sample, we see clear differences in the discs of FUor and EXor sources,  the former group being significantly more massive than the latter group. EXor objects also lack the prominent outflows seen in the FUor systems in our sample and in the borderline FUor object V1647 Ori. These differences suggest that FUor objects  represent an earlier stage in the disc evolution process than EXor sources.
 \\

\noindent 2) The inferred disc masses for the three FUor objects are $\sim$0.1--0.6 M$_{\odot}$, implying GI could play a role in their outbursts. 
On the other hand, the inferred disc masses for the faintest EXor targets are $\lesssim$ 1-5 M$_{JUP}$, and  thus alternative mechanisms must be responsible for their outbursts. \\

\noindent 3) The discs around FUor objects are compact  (R$_c$ $\sim$20--40 au).  They have smaller radii for a given disc mass and are more centrally concentrated than T Tauri discs. This could be an evolutionary effect and/or an intrinsic property of FUor objects. \\

\noindent 4)  V1118 Ori, the only known close binary system in our sample ($\sim$75 au separation), is clearly resolved into two millimeter sources, indicating the presence of a disc around each of the stellar components.  
The disc around HBC~494  is asymmetric, suggesting structure in the outer disc or a stellar companion also surrounded by a disc.

\section*{Acknowledgments}
This paper makes use of the following ALMA data: ADS/JAO.ALMA \#2013.0.00710.S.  ALMA is a partnership of ESO (representing its member states), NSF (USA) and NINS (Japan), together with NRC (Canada), NSC and ASIAA (Taiwan), and KASI (Republic of Korea), in cooperation with the Republic of Chile. The Joint ALMA Observatory is operated by ESO, AUI/NRAO and NAOJ.
The National Radio Astronomy Observatory is a facility of the National Science Foundation operated under cooperative agreement by Associated Universities, Inc.
L.A.C. was supported by CONICYT-FONDECYT grant number 1171246. L.A.C., S.C. and A.Z. acknowledge support from the Millennium Science Initiative (Chilean Ministry of Economy),  through grant ``Nucleus RC130007''.  Support for JLP is provided in part by FONDECYT through the grant 1151445 and by the Ministry of Economy, Development, and TourismÕs Millennium Science Initiative through grant IC120009, awarded to The Millennium Institute of Astrophysics, MAS.
The modelling and MCMC calculations used in this work were performed
in the Belka and Strelka clusters, financed by Fondequip project
EQM140101 and housed at MAD/Universidad de Chile/Cerro Calan.

\vspace{0.5cm}


\bsp

\label{lastpage}

\end{document}